# The effects of degree correlations on network topologies and robustness [*]

Zhao Jing[1,2,4], Tao Lin[2], Yu Hong[2], Luo Jian-Hua[1], Cao Zhi-Wei[2§] and Li Yi-Xue[2,3,1§]

[1]School of Life Sciences & Technology, Shanghai Jiao Tong University, Shanghai 200240, China
[2]Shanghai Center for Bioinformation and Technology, Shanghai 200235, China
[3]Shanghai Institutes for Biological Sciences, Chinese Academy of Sciences, Shanghai 200031, China.
[4]Department of Mathematics, Logistical Engineering University, Chongqing 400016, China.

**Abstract.** Complex networks have been applied to model numerous interactive nonlinear systems in the real world. Knowledge about network topology is crucial for understanding the function, performance and evolution of complex systems. In the last few years, many network metrics and models have been proposed to illuminate the network topology, dynamics and evolution. Since these network metrics and models derive from a wide range of studies, a systematic study is required to investigate the correlations between them. The present paper explores the effect of degree correlation on the other network metrics through studying an ensemble of graphs where the degree sequence (set of degrees) is fixed. We show that to some extent, the characteristic path length, clustering coefficient, modular extent and robustness of networks are directly influenced by the degree correlation.

## 1. Introduction

Complex networks have been playing an important role in the understanding of many natural systems[1-6]. The empirical analysis of many real-world networks from various realms has revealed some common features shared by the individual systems, such as power-law degree distribution[7], modular or community organization[8-11], small-world effect [12], and hierarchical organization [13], as well as some apparent differences between networks, such as the positive degree correlation in social networks vs. the negative degree correlation in technological and biological networks[14-19]. At the same time, several network metrics has been proposed to describe network topology quantitatively, in which the most important and commonly used ones are the degree distribution [7], the characteristic path length [12], clustering coefficient [12], modularity[10, 20], and assortative coefficient[15]. On the other hand, a considerable number of network models have been developed to simulate networks in the real world, in which the most widely used ones are the BA preferential attachment model for the evolution of networks[7], and the RB model for the hierarchically modular organization pattern of networks[21]. Since the network metrics and models have derived from a wide range of studies, a systematic study is required to investigate the possible correlations between the network metrics, as well as the applicability of network models.

---

[*] This work was supported in part by grants from Ministry of Science and Technology China(2003CB715900, 04BA711A21, 2004CB720103)，National Natural Science Foundation of China (30500107)，and Science and technology commission of Shanghai municipality（04DZ19850, 04DZ14005）.

§ Corresponding author: zwcao@scbit.org; yxli@sibs.ac.cn.

In this paper, we study how the network topologies and performances are influenced by the degree correlation through investigating a statistical ensemble generated from a certain "seed" network, which includes the maximally random networks under the constraints of given degree sequence as well as fixed assortativity [22]. Two model networks and two real biological networks are applied as seed networks.

## 2. Seed Networks

In this work, four networks whose degree distributions obey the power-law are studied as seed networks:

(1) Seed network A: the hierarchically modular network constructed by Ravasz et al. (RB model) [21] in the 3rd iteration. At every step, the central node of the central module is connected to the external nodes of the four peripheral modules (see figure 1 of [21]).
(2) Seed network B: a model network constructed by the BA preferential attachment model with parameter values m = m0 = 3 [7].
(3) Seed network C: the biggest connected cluster of the *E.coli* metabolic network[8], in which nodes correspond to substrates and edges to reactions. In order to reflect biologically relevant transformations of substrates, the top 10 most common substrates are removed[8, 23].
(4) Seed network D: the biggest connected cluster of the protein interaction network CCSB-HI1, a proteome-scale map of human binary protein–protein interactions tested from a high-throughput yeast two-hybrid experimental system [24].

In table 1, we summarize the basic graph metrics of these networks. We quantify the topological features of the networks by their characteristic path length (L), clustering coefficient (C) and modularity (M), and the degree correlation by assortative coefficient (r).

Table 1. Basic graph metrics of the seed networks.

| Seed network | A | B | C | D |
|---|---|---|---|---|
| Number of nodes | 625 | 1000 | 716 | 1307 |
| Number of links | 1976 | 2991 | 1425 | 2483 |
| Degree distribution | $P(k) \sim k^{-2.42}$ | $P(k) \sim k^{-3}$ | $P(k) \sim k^{-2.78}$ | $P(k) \sim k^{-2.51}$ |
| Characteristic path length (L) | 3.36 | 3.59 | 4.54 | 4.36 |
| Clustering Coefficient (C) | 0.5821 | 0.0246 | 0.0928 | 0.0327 |
| Modularity (M) | 0.688 | 0.4046 | 0.6637 | 0.587 |
| Assortative coefficient (r) | -0.1516 | -0.0527 | -0.084 | -0.2228 |

## 3. Network ensemble based on the spectrum of degree-correlation

*3.1. Topological metrics of degree correlation*

The degree correlation of a network is the pattern of correlations between degrees of neighboring nodes [14-19]. It could be simply measured by the assortative coefficient[15], defined as the Pearson's correlation coefficient of degrees at either side of an edge, whose theoretical range is [-1, 1]. High values of assortative coefficient mean the tendency for high-degree nodes to attach to other high-degree ones, and low values represent the preference of linkages between high-degree nodes and low-degree ones.

There are some other ways of quantifying degree correlations[18, 25, 26]. Li et al. [26] defined the *s*-metric as the sum of products of degrees of adjacent nodes as follows:

$$s = \sum_{(i,j) \in E} d_i d_j \quad (1)$$

where $d_i$ is the degree of node *i*, $E$ is the edge set of the graph. They also proved that *s*-metric is linearly correlated to the assortativity coefficient[26]. Therefore, for a given degree sequence, the biggest and smallest *s*-value, denoted by $S_{max}$ and $S_{min}$ respectively, correspond to the extremes of assortativity. The simple connected graphs with the extreme *s*-values are called $S_{max}$ graph and $S_{min}$ graph respectively.

*3.2. Generating extreme networks of degree correlation*

To construct the ensemble of networks, we first generated the extreme graphs of degree correlation from the seed network. They are networks that have the highest and lowest value of assortative coefficient while conserving the degree sequence of the seed network.

We apply the algorithm proposed by Li *et al.* to construct the extreme graphs[26]. This algorithm uses an ordered sequence of all potential links as its input, and then checks the sequence from top to bottom to select one edge at a time so as to result in a simple, connected graph of given degree sequence *D*. Ref [26] constructs the $S_{max}$ graph for a given seed network, in which all potential links (*i*, *j*) for all node pairs *i*, *j* ($i<j$, $d_i \geq d_j$; *i*, *j*=1,2,…*N*) are ordered according to their weight $d_i d_j$ in a decreasing way.

Here, we first order the potential links for the $S_{max}$ graph and $S_{min}$ graph according to the Rearrangement Inequality[26]:

$$a_1 b_1 + a_2 b_2 + \ldots + a_n b_n \geq a_1' b_1 + a_2' b_2 + \ldots + a_n' b_n \geq a_n b_1 + a_n b_2 + \ldots + a_n b_n,$$

where $a_1 \geq a_2 \geq \ldots \geq a_n$, $b_1 \geq b_2 \geq \ldots \geq b_n$, $(a_1', a_2', \ldots a_n')$ is any permutation of $(a_1, a_2, \ldots, a_n)$. Then we apply the ordered sequences as inputs of the algorithm in ref [26] to construct the $S_{max}$ graph and $S_{min}$ graph, respectively.

Obviously, there are a lot of permutations of potential links that satisfy the Rearrange Inequality, especially when the degree sequence includes many nodes of the same degrees. We order the potential links as follows:
1. For the degree sequence of the seed network, order the nodes according to decreasing values of node degree, i.e., $d_1 \geq d_2 \geq \ldots \geq d_N$. More specifically speaking, the nodes are labeled as $(1,2,\ldots k_1, k_1+1,\ldots,k_2,\ldots,k_{m-1}+1,\ldots,k_m)$ such that

$$d_1 = d_2 = \ldots = d_{k_1} > d_{k_1+1} = \ldots = d_{k_2} > \ldots > d_{k_{m-1}+1} = \ldots = d_{k_m}, \quad k_m = N.$$

2. For each node $i$ ($i=1, 2, \ldots, N-1$), order its potential links for the $S_{max}$ ($S_{min}$) graph decreasingly (increasingly) according to the products of degrees of the nodes at either side of an edge.

(a) For the $S_{max}$ graph, order all potential links of node $i$ as follows:
$$L_i = \{(i,i+1),(i,i+2),\ldots(i,N)\}.$$

(b) For the $S_{min}$ graph, suppose $k_{p-1} \leq i < k_p$, order all potential links of node $i$ as follows:

$$L_i' = \begin{cases} \{(i,k_{m-1}+1),\ldots,(i,N),(i,k_{m-2}+1),\ldots,(i,k_{m-1}),\ldots,(i,i+1),\ldots(i,k_p)\}, & \text{if } p < m \\ \{(i,i+1),(i,i+2),\ldots(i,N)\} & \text{if } p = m \end{cases}$$

3. (a) For case 2(a), order all potential links of the $S_{max}$ graph as $L_1, L_2, \ldots, L_{N-1}$.

(b) For case 2(b), order all potential links of the $S_{min}$ graph as $L_1', L_2', \ldots, L_{N-1}'$.

In Figure 1 we show the resulting extreme graphs for a small seed network as an example. As can be seen, the extreme networks generated by this algorithm exhibit a chain-like topology. In this $S_{max}$ graph, nodes with similar degrees are linked to form quasi-cliques, while in the $S_{min}$ graph, high degree nodes are matched with low degree nodes to form bipartite quasi-cliques, and the remained unmatched nodes are linked with each other in a clique-like modular way.

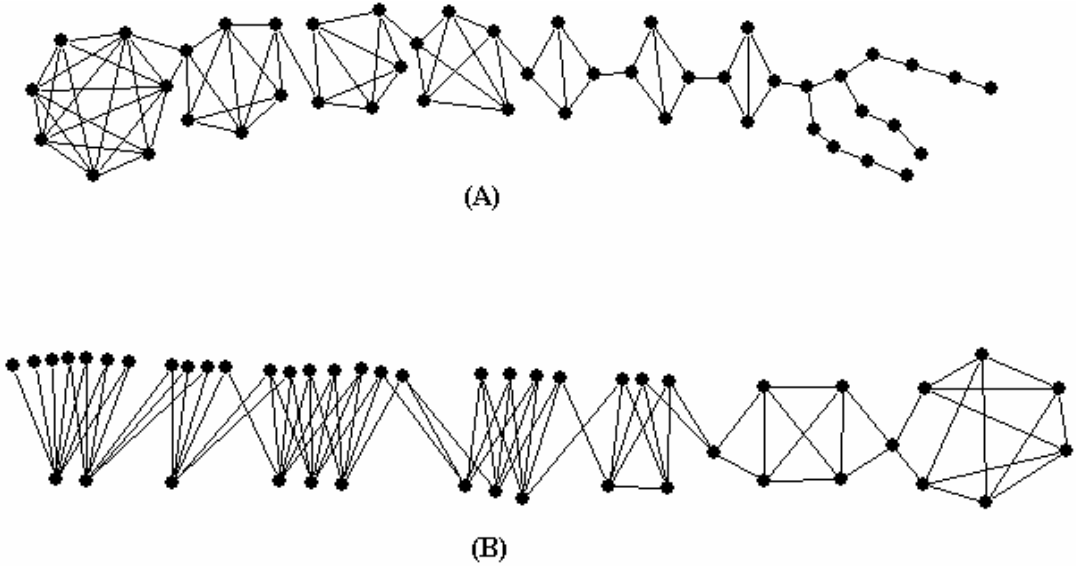

Fig. 1. The $S_{max}$ graph (A) and $S_{min}$ graph (B) for a small seed network.

*3.3 Constructing network ensemble from the $S_{max}$ and $S_{min}$ graphs*

Starting from the $S_{max}$ and $S_{min}$ graph respectively, we carried out a series of edge rewiring in two steps as follows:

Step 1: In one direction, we rewire the links of $S_{max}$ and $S_{min}$ graph respectively, during which two links between two randomly selected pairs of nodes are rewired only if this step generates no multiple edges while keeps the graph connected[19, 25].

The process was repeated $pM$ times in which only performed exchanges were counted, where $M$ is the total number of links in the graph, and $p$ is a positive number representing the rewiring fraction of links. Switching $p$ in the area [0, 3], we generated a set of randomized networks for the $S_{max}$ graph and the $S_{min}$ graph, respectively. For $p=0$, the original $S_{max}$ and $S_{min}$ graph are unchanged.

Step 2: In another direction, we rewire each network got in step 1 in a way that keeps its degree sequence and assortative coefficient. That is to say, a pair of randomly selected edges ($v_1$, $v_2$) and ($v_3$, $v_4$) is rewired to ($v_1$, $v_3$), ($v_2$, $v_4$) only if they satisfy the following conditions:

(1) Edges ($v_1$, $v_3$), ($v_2$, $v_4$) do not exist in the original graph
(2) $d(v_1)*d(v_2)+d(v_3)*d(v_4) = d(v_1)*d(v_3)+d(v_2)*d(v_4)$
(3) The new graph is still connected

This process was repeated sufficiently large number of times. Thus every network generated from step 2 is the maximally random network under the constraints of given degree sequence as well as fixed assortativity [22]. All of these resulting networks constitute the network ensemble in this study, and we denote the ensemble of seed network G as $\mathcal{G}_D^A(G)$.

## 4. Numerical results

In this section, we will study the possible relation between degree correlation and the other structural metrics numerically by the resulting network ensemble $\mathcal{G}_D^A(G)$ of our four seed networks.

*4.1 Assortative coefficient*

To probe the spectrum of degree correlation in each ensemble, we plot the assortative coefficient $r$ of the networks generated in step 1 as a function of the rewiring fraction $p$ in Figure 2. It can be seen that the change tendencies of assortative coefficient $r$ for the randomization of the $S_{max}$ graph and the $S_{min}$ graph are opposite. As $p$ increases from 0 to 1, assortative coefficients change according between the maximum and minimum values, and then they are convergent to a constant when $p \geq 1$. This tendency suggests that after performing more than $M$ times ($p \geq 1$), the random rewiring process reaches a good mixing, thus generates the maximally random networks under only one constraint of given degree sequence [22]. We denote this maximally random network ensemble of seed network G as $\mathcal{G}_D(G)$, and the limit value of the assortative coefficient as $r_c$. Hence the latter corresponds to the average $r$-value of the $\mathcal{G}_D(G)$ ensemble. It is worth to note that both the ensemble $\mathcal{G}_D(G)$ and $\mathcal{G}_D^A(G)$ represent the statistical sets of randomized networks with given degree sequence, yet the randomized extents of networks in these ensembles are different. In the process of randomization, networks in $\mathcal{G}_D^A(G)$ have one more constraint of assortativity than those in $\mathcal{G}_D(G)$.

Figure 2 indicates that the rewiring process in step 1 allows us to tune the network between the maximum and minimum level of degree correlation, and the resulting network set spans the whole range of assortativity based on the same degree sequence. Therefore, the ensemble $\mathcal{G}_D^A(G)$, which is constructed by a randomization process which keeps the degree sequence and assortativity as described in step 2, could be used to probe the transformation of network topologies depending on that of its degree correlation.

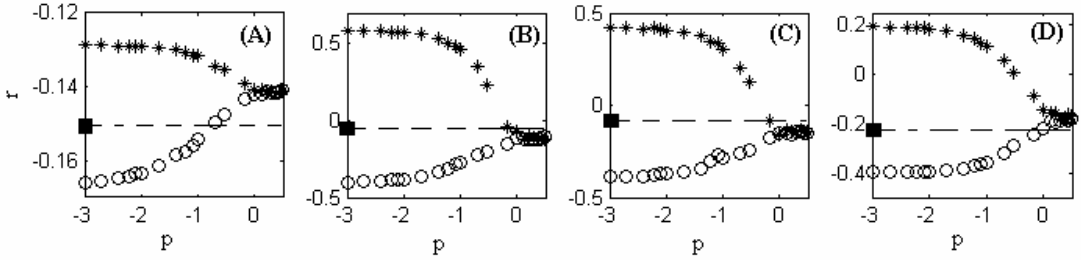

Fig.2. Assortative coefficient (*r*) as function of the randomization fraction (*p*). A logarithmic horizontal scale is used to depict the mini increasing of rewiring fraction. The square corresponds to the (*p*, *r*)-value of the seed network. The stars and circles indicate the values of the randomized networks from the $S_{max}$ graph and the $S_{min}$ graph, respectively.

Figure 2 also illustrates that the practice ranges of assortative coefficient are far less than the theoretical limit [-1, 1], and not all the maximally random networks under the constraint of given degree sequence ($p \geq 1$) have assortative coefficient near 0. Except those of seed network B, maximally random networks for the other seed networks are disassortative mixing with negative assortative coefficient. This observation could be explained by the structural cut-off constraints, i.e., simple random graphs whose degree distribution follows a power law with the exponent $\gamma$ must satisfy the condition of $\Delta \leq \sqrt{N}$ ($\Delta$ and $N$ are the highest degree and total number of nodes respectively) in order to have no degree-correlations[27]. This cut-off constraint is equivalent to $\gamma \geq 3$ [27, 28], which is only satisfied by network B. Figure 2 also displays that even the assortative coefficient of the $S_{max}$ graph for seed network A is negative, suggesting that any connected simple graph having the same degree sequence as network A is disassortative mixing, no matter how its nodes are connected. This could be caused by the highly heterogeneous distribution of node degree in network A, in which $\Delta > \frac{1}{2}N$ but the total number of high-degree nodes is much less than $\frac{1}{2}N$. Thus the high-degree nodes have to link with much more low-degree nodes than high-degree ones, resulting in the disassortative mixing feature.

## 4.2 *Characteristic path length*

Characteristic path length is defined as the average length of the shortest paths between any pair of nodes in the network[12]. It can measure the average communication speed between the nodes of the network.

As can be seen from Figure 1, the extreme networks are linked in a chain-like way. It is true that all of the resulting extreme networks of the four seed networks, except the $S_{max}$ graph for seed network D, have chain-like structures with very long paths. That is to say, the characteristic path lengths of these networks are much larger than those in logarithmical scale of the network size. The reason of the exception (the $S_{max}$ graph for seed network D) is that network D has a lot of degree-one nodes that need to link with high-degree nodes for keeping the connectivity of the network. Although the characteristic path length of almost all scale-free networks in the real world scale logarithmically with the system size (called small-world networks), our result suggests that scale-free networks could also be large-world networks. Actually, network models with power-law degree distribution but large characteristic path length have been proposed in [29-31].

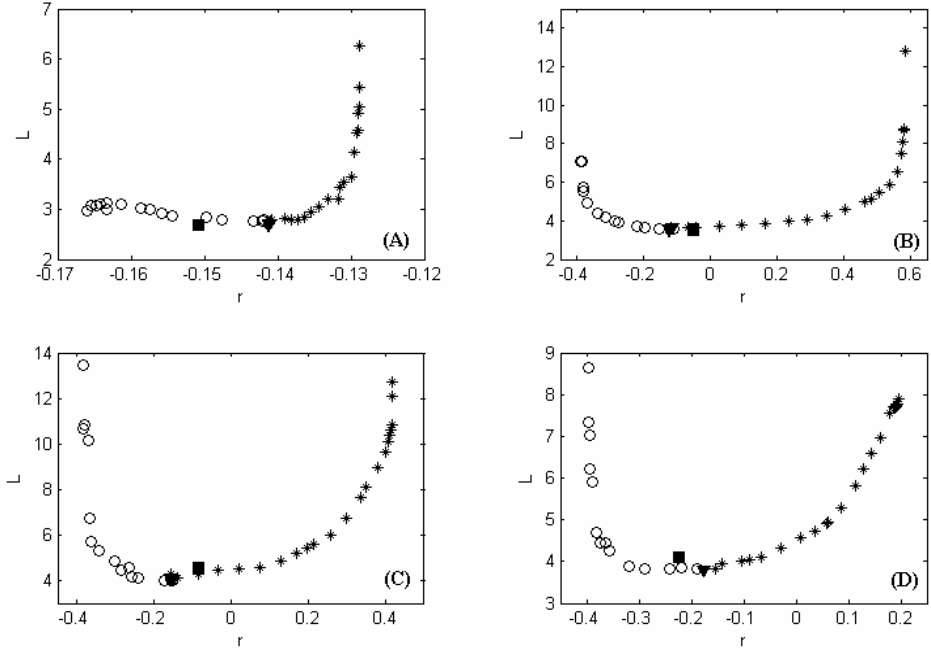

Fig.3. Relationship between characteristic path length (L) and assortative coefficient (r). The triangle corresponds to the average (r, L)-value of the $\mathcal{G}_D(G)$ ensemble, and the other symbols correspond to those of Figure 2. The data shown in the figures are averaged over 10 random realizations of the rewiring process.

Figure 3 displays the relationship between the characteristic path length (L) and assortative coefficient (r) in the network ensemble $\mathcal{G}_D^A(G)$. As can be seen, the characteristic path length of the seed networks lies close to the minimum value, which is also the average *L*-value in the $\mathcal{G}_D(G)$ ensemble. The characteristic path length approaches the minimum when *r* is near to $r_c$, and it decreases before $r = r_c$ and increases after $r = r_c$.

*4.3 Clustering coefficient*

It has been reported that many categories of real-world networks are enriched with triangles. In friendship networks, the high density of triangles could be caused by the fact that friends of the same people are usually friends of each other. Clustering coefficient is a network metric to quantify the density of triangles in a network. Here we applied the definition of ref [12], in which the clustering coefficient of node *v* is defined as the probability that two neighbor nodes of *v* are also connected with each other:

$$CC(v) = \frac{2|N(v)|}{d(v)(d(v)-1)}, \qquad (2)$$

where $|N(v)|$ denotes the number of links between neighbors of node *v*, *d(v)* is the degree of node *v*. The clustering coefficient of a network is defined as the average of *CC (v)* over all *v*.

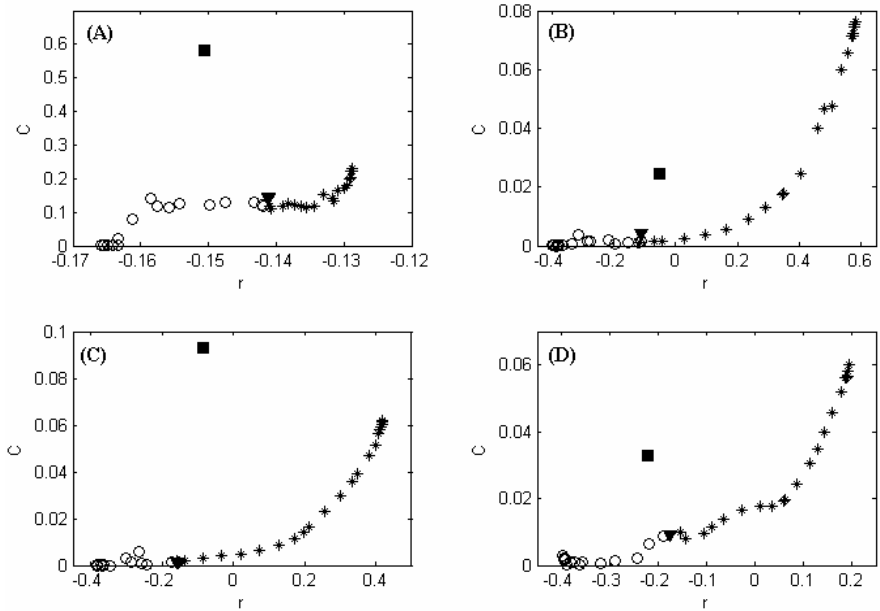

Fig.4. Relationship between clustering coefficient(C) and assortative coefficient (r). The symbols correspond to those of Figure 3. The data shown in the figure are averaged over 10 random realizations of the rewiring process.

Figure 4 shows that clustering coefficient is a monotonously increasing function of assortative coefficient. This result could be explained intuitively from Figure 1: networks with higher assortativity include more quasi-cliques, in which there are high densities of triangles; whereas those with lower assortativity contain more bipartite sub-graphs, in which there are no triangles. The correlation between these two variables also agrees with the earlier observation [32]. In addition, Figure 4 displays that the seed network is significantly more clustered than the

$\mathcal{G}_D(G)$ average.

*4.4 Modularity*

Many real-world networks, including social networks, technological networks, and biological networks, are found to divide naturally into communities or modules, i.e., densely connected sub-networks within which there is a high density of edges, and between which there is a lower density of edges[10]. The modularity metric was initially defined to quantitatively compare different partitions of a network [10]:

$$M = \sum_{i=1}^{r}[e_{ii} - (\sum_{j} e_{ij})^2] \qquad (3)$$

where $r$ is the number of clusters, $e_{ij}$ is the fraction of edges that leads between vertices of cluster $i$ and $j$. Guimera et al. then defined the modularity metric of a network as the largest modularity metric of all possible partitions of the network[20], and developed a simulated annealing algorithm to compute this metric[33, 34]. Thus the modularity metrics of networks could be applied to measure the modular extent of different networks.

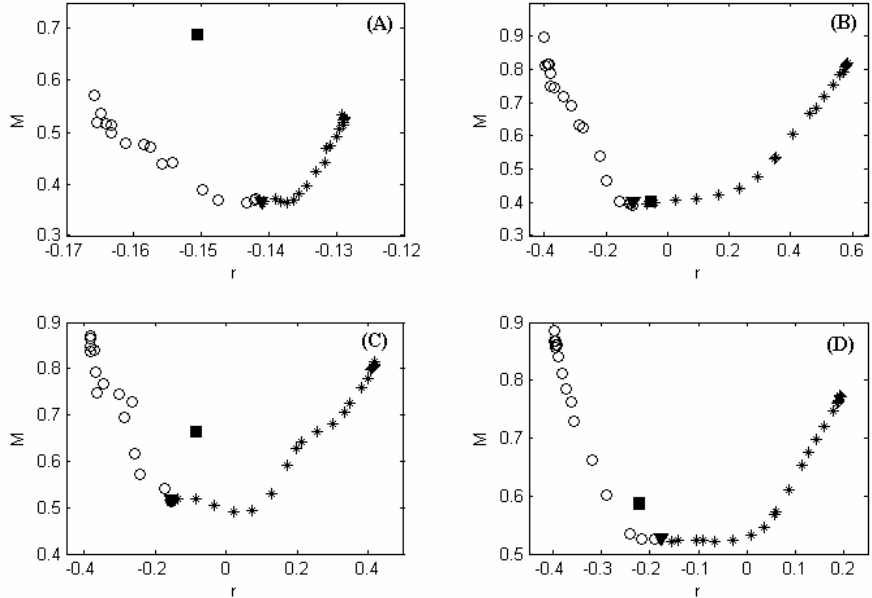

Fig.5. Relationship between modularity($M$) and assortative coefficient ($r$). The symbols correspond to those of Figure 3. The data shown in the figures are averaged over 10 random realizations of the rewiring process.

Figure 5 depicts the modularity as a function of the assortativity, illustrating that networks with the assortative coefficient near to $r_c$ exhibit the lowest extent of modularity, and the closer the assortative coefficient $r$ of a network is to the maximum or minimum value, the bigger its modularity $M$ is. Newman and Park have suggested that the formation of communities (or modules) could produce highly assortative mixing[17]. Our results indicate that the significantly high extent of assortative mixing or disassortative mixing could result in high level of modularity. The phenomenon that the networks of extreme assortativity exhibit high extent of modularity could be observed directly from the extreme networks shown in Figure 1. Actually, networks that show both rich and varied community structure as well as disassortative mixing has also been found in some network models of mechanical assemblies [35].

Figure 5 also shows that the real biological networks, seed network C and D, has significantly higher extent of modularity than networks with the same assortativity in the ensemble $\mathcal{G}_D^A(G)$, suggesting the modularity might be a intrinsic feature of biological networks developed during the long evolutionary process. However, seed network B, a realization of BA model, does not exhibit significantly high level of modularity, indicating that the BA model could not simulate the modular feature of real networks as the RB model does.

*4.5 Network resilience*

Most dynamical processes taking place in the topology defined by the network, such as information communication and disease spreading, rely heavily on the connectivity of networks. When some nodes are removed, the network might be fragmentized and the interactions between nodes are thus interrupted. Along this line, two ways of damages to networks - intentional attacks targeting hubs (attacks) and random removal of nodes (failures), as well as the network performances due to these actions have been usually studied[36].

In this study, we measure the network performance under attacks or failures by the *f-robustness* metric, defined as the fraction of vertices expected to be removed so that the ratio between the size of the largest connected component for the resulting network and that of the original network would equal to $f \in (0, 1)$[37]. We set $f = 1/2$, and denote the 1/2-rubustness under attacks and failures as $R_a$ and $R_f$, respectively. In Figure 6, we display $R_a$ and $R_f$ respectively as a function of assortativity. Comparison of the plots in the first row with the corresponding ones in the second row shows that, for the same network, the value of $R_f$ is significantly bigger than that of $R_a$. This observation agrees with the earlier finding that scale-free networks are resilient to random failures but relatively vulnerable to intentional attacks[36].

The first row of Figure 6 illustrates that the robustness under attacks is roughly an increasing function of assortativity. The only exception is the area of the assortative coefficient near the maximum value, in which $R_a$ decreases with assortativity. This transition may stem from the chain-like topologies of these networks. Although $R_a$ does not monotonously increases in the area near the maximum assortativity, Figure 6 still suggests that, compared with dissortative networks, assortative networks are more robust to removal of their highest degree nodes, supporting the argument in [15, 38]. The second row of Figure 6 shows that networks with assortative coefficient near $r_c$ exhibit the highest extent of robustness under failures, and the robustness decreases with the approaching of the assortative coefficient to the extreme values.

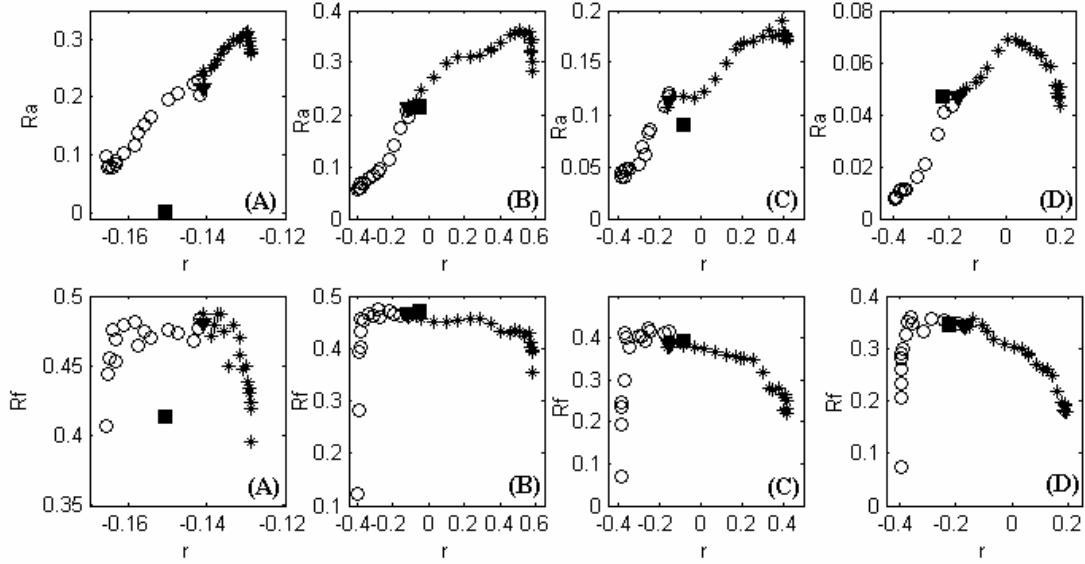

Fig. 6. The effect of degree correlation on network robustness. The symbols correspond to those of Figure 3. Figures in the first and second row depict the robustness under attacks and failures as a function of assortativity, respectively. The data shown in the figures are averaged over 10 random realizations of the rewiring process.

As can be seen, the two real biological networks, seed network C and D, exhibit almost the best performance under failures in their own ensembles, so does seed network B, a network constructed by the BA model. These three networks also show nearly the same extent of robustness under attacks as the networks with the same assortativity in the ensemble $G_D^A(G)$. Thus the higher extent of robustness of real biological networks might be a result of evolution. However, although seed network A has been used as a model to characterize real-world networks such as metabolic networks and protein interaction networks[21], its performances under attacks and failures are significantly worse than those of the real biological networks, suggesting that the RB model could not appropriately simulate the robustness of real networks as the BA model does.

## 5. Conclusion

In this study, through an exploring to the maximally random ensemble of networks under the constraints of given degree sequence and fixed assortativity, we have conducted a system-level survey about how the degree correlation of networks is related to its topologies and performances. As illustrated, this ensemble could afford us more information than before. By investigating this ensemble, we found that networks with the same power-law degree sequence could have very different topological features and performances under attacks and failures, and these features are partially determined by the degree correlation. In addition, our study indicates that both BA and RB model could simulate the small-world feature of real biological networks, i.e., short path length and high clustering. However, BA model is good at simulating the robustness but weak at mimicking the modular feature of real networks, whereas RB model is just opposite, suggesting that these models still have limitations in depicting the main features of real networks. Thus more network models are expected to be developed. The approach used in this study, which compare real networks with its random counterparts at different extents, places the network under study in a context of comparable statistical weight. This approach may help us to get clearer insight about the topological properties of the network, thus could be promising in complex network study.

**Acknowledgements**

We thank Dr. L. A. N. Amaral and Dr. R. Guimerà for kindly providing us the software Modul-w to compute network modularity metric; Dr. J. Doyle, Dr. P. Holme and Dr. L. Li for stimulating discussions and constructive comments.